\documentclass[pra,showpacs,amsfonts,amsmath,amssymb,twocolumn,a4paper]{revtex4}
\usepackage{natbib}
\usepackage{relsize}
\usepackage[T1]{fontenc}
\usepackage{ae}
\usepackage{color}
\usepackage{graphicx}
\usepackage{bbm}

\begin{document}

\title{Microwave-Induced Amplitude and Phase Tunable Qubit-Resonator Coupling\\in Circuit Quantum Electrodynamics}

\author{S.~Zeytino$\breve{\mathrm{g}}$lu\footnote{These authors contributed equally to this work.}}
\author{M.~Pechal\footnotemark[\value{footnote}]}
\email{mpechal@phys.ethz.ch}
\author{S.~Berger}
\author{A.~A.~Abdumalikov Jr.}
\author{A.~Wallraff}
\author{S.~Filipp\footnote{Now at IBM T.J. Watson Research Center, Yorktown Heights, NY 10598, United States}}
\address{Department of Physics, ETH
Zurich, CH-8093
  Zurich, Switzerland}
\date{\today}
\pacs{42.50.Ct, 03.67.Lx, 42.50.Pq, 85.35.Gv}

\renewcommand{\i}{{\mathrm i}}
\def\1{\mathchoice{\rm 1\mskip-4.2mu l}{\rm 1\mskip-4.2mu l}{\rm 1\mskip-4.6mu l}{\rm 1\mskip-5.2mu l}}
\newcommand{\ket}[1]{|#1\rangle}
\newcommand{\bra}[1]{\langle #1|}
\newcommand{\braket}[2]{\langle #1|#2\rangle}
\newcommand{\ketbra}[2]{|#1\rangle\langle#2|}
\newcommand{\opelem}[3]{\langle #1|#2|#3\rangle}
\newcommand{\projection}[1]{|#1\rangle\langle#1|}
\newcommand{\scalar}[1]{\langle #1|#1\rangle}
\newcommand{\op}[1]{\hat{#1}}
\newcommand{\vect}[1]{\boldsymbol{#1}}
\newcommand{\id}{\text{id}}
\newcommand{\red}[1]{\textcolor{red}{#1} }

\begin{abstract}
In the circuit quantum electrodynamics architecture, both the resonance frequency and the coupling of superconducting qubits to microwave field modes can be controlled via external electric and magnetic fields to explore qubit -- photon dynamics in a wide parameter range. Here, we experimentally demonstrate and analyze a scheme for tuning the coupling between a transmon qubit and a microwave resonator using a single coherent drive tone. We treat the transmon as a three-level system with the qubit subspace defined by the ground and the second excited states. If the drive frequency matches the difference between the resonator and the qubit frequency, a Jaynes-Cummings type interaction is induced, which is tunable both in amplitude and phase. We show that coupling strengths of about $10~\rm{MHz}$ can be achieved in our setup, limited only by the anharmonicity of the transmon qubit. This scheme has been successfully used to generate microwave photons with controlled temporal shape [Pechal {\it et al.}, Phys. Rev. X {\bf 4}, 041010 (2014)]
and can be directly implemented with superconducting quantum devices featuring larger anharmonicity for higher coupling strengths.
\end{abstract}

\maketitle

\section{Introduction}

The strength of the interaction between an atom and the electromagnetic field is determined by the dipole
moment of the atom and the mode volume of the electromagnetic
field. Although in free space the interaction is typically weak, it can
be enhanced by confining the field to a small volume in a cavity quantum electrodynamics (QED) setting. This field has seen tremendous progress~\cite{Thompson1992,Miller2005,Haroche2006} and
has diversified from the traditional setting with real atoms to solid
state realizations using nanoscale electronic devices such as quantum dots \cite{Reithmaier2004,Khitrova2006} or superconducting
circuits \cite{Wallraff2004,Chiorescu2004,Schoelkopf2008} as artificial atoms. However, in most of the solid-state settings the coupling strength between the atom and the cavity modes is fixed by the
geometry of the device and the position of the artificial atom in the cavity, neither of which
can be modified $\textit{in situ}$. Although in recent years, superconducting circuit devices
which allow $\textit{in situ}$ access to the amplitude of the qubit-resonator coupling have been realized
\cite{Hime2006,Bialczak2011,Gambetta2011,Srinivasan2011}, a scheme for controlling the phase of this coupling has only recently been demonstrated in \cite{Pechal2014}. Such a scheme is expected to be useful in a variety of settings, such as quantum
gate operations \cite{Romero2012}, creating shaped photons for
quantum networks \cite{Cirac1997,Kuhn2002,Pechal2014}, measuring the vacuum state of a cavity \cite{Oi2013}, exploring
vacuum-induced Berry phases \cite{Fuentes-Guridi2002}, enabling the controlled coupling of a single or multiple qubits to multiple resonator modes \cite{Larson2011,Bose2003a,Larson2012}, or engineering quantum reservoirs \cite{Poyatos1996}.

The amplitude and phase tunability of the qubit-resonator coupling strength can be achieved by a two-photon process, known as a
cavity-assisted Raman process, which has been extensively studied
for $\Lambda$-systems \cite{Gerry1990,Cardimona1991,Alexanian1995,Wu1996,Law1997}
and ladder-type ($\Xi$) systems \cite{Wu1997}. Cavity-assisted Raman processes employ an external coherent drive with a time-dependent amplitude $\Omega\cos{(\omega_d t+\phi)}$ to induce an effective coupling $\tilde{g}$ between the qubit and the resonator degrees of freedom. The experimental access to the amplitude ($\Omega$) and the phase ($\phi$) of the external drive enables $\textit{in situ}$ amplitude and phase tunability of this effective qubit-resonator coupling.

Cavity-assisted Raman processes can be readily applied to superconducting circuit elements of the transmon-type \cite{Koch2007}, which
are in wide-spread use because of their excellent coherence properties
and the relative simplicity of their fabrication. These circuit elements
realize an anharmonic oscillator system, i.e., a system in which transitions are allowed only between neighboring states and the transition frequencies differ from each other by multiples of a small negative parameter $\alpha$ which characterizes the anharmonicity. In experiments using the circuit QED architecture, not only the
transition between ground $\ket{g}$ and the first
excited $\ket{e}$ state at frequency $\omega_{ge}$, but also
transitions between higher lying energy levels
can easily be addressed \cite{Bianchetti2010} and complex quantum states can be realized \cite{Shalibo2013}. In particular, the second excited state
$\ket{f}$, which is separated from $\ket{e}$ by $\omega_{ef} = \omega_{ge} + \alpha$, has been used widely for
quantum gates~\cite{Strauch2003,DiCarlo2009,Fedorov2012,Abdumalikov2013}, and plays an important role in our implementation of the cavity-assisted Raman processes in a circuit QED setting.

In our experiments, we investigate the tunability of a cavity-assisted Raman process induced coupling between a microwave resonator and a transmon device whose qubit states are defined as the ground and second excited states. We demonstrate the amplitude tunability of the transmon-resonator coupling, and analyze the effects of the small anharmonicity of the transmon on the rate and fidelity of the population exchange (swap) between the transmon and the resonator, thereby complementing our experiments in which shaped microwave photons have been created and analyzed \cite{Pechal2014}.

The outline of the paper is as follows. In Section~$\mathrm{\ref{sec:spectroscopy}}$ we present spectroscopic measurements of the tunable coupling strength $\tilde{g}$ using a transmon-type superconducting qubit. In Section~$\mathrm{\ref{sec:method}}$, we derive an analytical expression for $\tilde{g}$ using first order perturbation theory in the drive amplitude $\Omega$. In Section~$\mathrm{\ref{sec:ACStark}}$, we explain an iterative method for calculating the drive-induced ac Stark shift of the qubit levels. In Section~$\mathrm{\ref{sec:numerical}}$, we use numerical simulations and second order perturbation theory to analyze the fidelity of the excitation exchange between the transmon and the resonator induced by the tunable coupling.

\section{Spectroscopic measurement of the tunable transmon-resonator coupling}
\label{sec:spectroscopy}

Our system consists of a transmon-type superconducting qubit with maximal Josephson coupling energy $E_{\mathrm{J}}^{\mathrm{max}}/h=47.3$~GHz and charging energy $E_{\mathrm{C}}/h=0.343$~GHz. We operate the transmon at a transition frequency of $\omega_{ge}/2\pi~=~8.103~\rm{GHz}$ which is higher than the fundamental mode frequency $\omega_r/2\pi~=~7.126~\rm{GHz}$ of the resonator, resulting in a positive transmon-resonator detuning $\Delta~=~\omega_{ge}-\omega_r~=~2\pi\times \,0.977~\rm{GHz}$. 
With a coupling strength $g/2\pi~=~65~\rm{MHz}$ between the transmon g-e transition and the fundamental resonator mode, the system is far in the dispersive regime ($\Delta~\gg~g$). The transmon has an anharmonicity $\alpha/2\pi~=~-0.376~\rm{GHz}$, and the frequency of the transition between the first and second excited state is $\omega_{ef}~=~\omega_{ge}~+~\alpha~=~2\pi\times7.727~\rm{GHz}$. The resonator decay rate $\kappa/2\pi$ is measured to be $6.6~\rm{MHz}$.

A tunable effective Jaynes-Cummings-like coupling between the transmon state $|f\rangle$ and the resonator can be activated by applying a coherent microwave tone at a frequency close to the energy difference between the dressed states $\ket{f0}_D$ and $\ket{g1}_D$. Here, the subscript $D$ denotes the dressing of the combined eigenstates $\ket{f0}\equiv \ket{f}\otimes\ket{0}$ and $\ket{g1}\equiv \ket{g}\otimes\ket{1}$ by the transmon-resonator coupling $g$. The effective coupling leads to a coherent excitation exchange between the transmon and the resonator. The fidelity of this exchange is maximum when the transmon drive frequency $\omega_d$ is equal to $\omega_{d}^{0}$, which is defined by the resonance condition
\begin{equation}
\omega_{d}^0=2\omega_{ge}+\alpha-\omega_{r}+\Delta_{f0g1}(\Omega).
\label{eq:resonantconditionstark}
\end{equation}
Thus, $\omega_{d}^{0}$ is the angular frequency difference between $\ket{f0}_D$ and $\ket{g1}_D$ modified by the difference $\Delta_{f0g1}(\Omega)$ between the ac Stark shifts for $\ket{f0}_D$ and $\ket{g1}_D$, which depends on the amplitude of the coherent drive $\Omega$.

\begin{figure}[ht!]
\centering
\includegraphics[width=85mm,trim=0mm 8mm 0mm 0mm, clip]{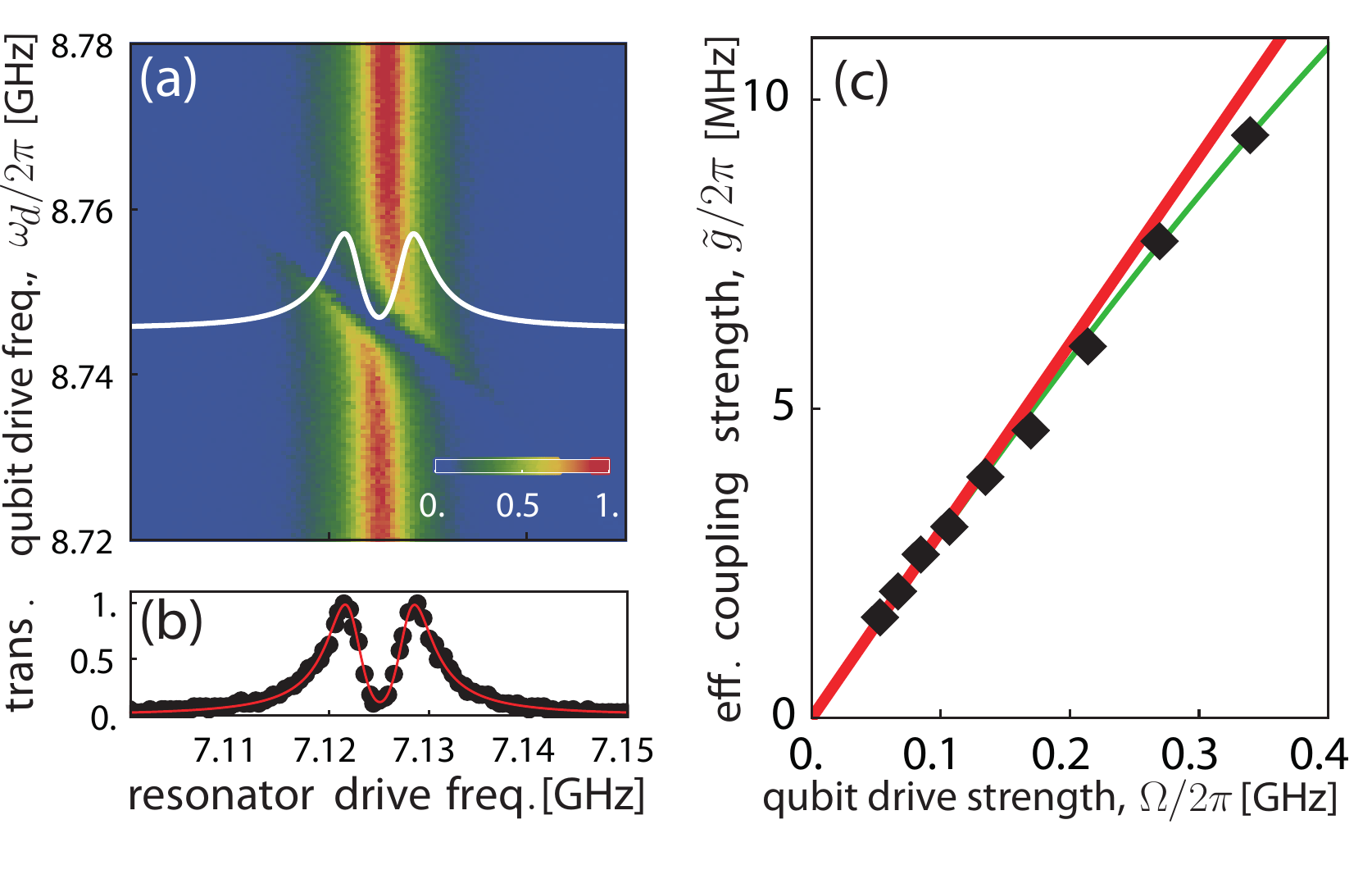}
\caption{(a) The resonator transmission as a function of transmon and resonator
  drive frequencies. The white curve is a fit to the data
  at the fixed transmon drive frequency $\omega_{d}/2\pi~=~8.75$~GHz. (b) The normalized transmission data for the white fit curve in (a). The amplitude of the effective coupling strength $\tilde{g}/2\pi$ extracted from this trace is $3.1~\rm{MHz}$. (c) The measured effective coupling $\tilde{g}$ (diamonds) as a function of coherent drive amplitude $\Omega$, the perturbation theory prediction (thick red line) and the numerical simulations (thin green line).}
\label{fig:splitting}
\end{figure}

The strength of the effective transmon-resonator coupling $\tilde{g}$ is measured by weakly probing the frequency-dependent transmission through the resonator \cite{Wallraff2004}. In the strong-coupling regime, the transmission peak of the resonator splits into two distinct peaks of equal width when the resonance condition in Eq. ($\ref{eq:resonantconditionstark}$) is satisfied, as observed in the measurement data shown in Fig.~$\ref{fig:splitting}$(a-b). The frequency separation between the two maxima in transmission then equals $2\tilde{g}/2\pi$, i.e., twice the coupling strength between the states $\ket{f0}_D$ and $\ket{g1}_D$.

To extract the effective coupling we first identify the transmon drive frequency $\omega_{d}^0$ at which the transmission curve is split into two peaks of identical width. We then fit the resonator transmission curve to the response function of the transmon-resonator system given by
\begin{eqnarray}
S(\omega_d) = A_{0}^{2}\left|\frac{i\left|\gamma\right|\omega-\tilde{\omega}^{2}}{4\tilde{g}^{2}(\omega_{d}^0)^{2}-(i\left|\gamma\right|\omega-\tilde{\omega}^{2})(i\left|\kappa\right|\omega-\tilde{\omega}^{2})}\right|^{2}
\label{eq:response}
\end{eqnarray}
derived from the master equation of the coupled transmon-resonator system in a truncated basis \cite{Carmichael1989}. In Eq.~($\ref{eq:response}$), $\gamma$ is the qubit decay rate, and $\tilde{\omega}^{2}=\omega^{2}-(\omega_{d}^{0})^{2}$.

We have measured the coupling strength $\tilde{g}$ for increasing amplitude of the coherent drive strength $\Omega$ [see~Fig.~$\ref{fig:splitting}$(c)]. When the drive is weak, the effective coupling strength $\tilde{g}$ increases linearly with the amplitude of the drive strength, in good agreement with a first-order perturbation theory calculation outlined in section \ref{sec:method}. However, for drive amplitudes larger than approximately $0.2~\rm{GHz}$, higher-order effects start to contribute significantly and the dependence of $\tilde{g}$ on $\Omega$ becomes non-linear, making the $\tilde{g}$ smaller than predicted by the linear model. The measured strength of the tunable coupling shows good agreement with the numerical simulation [Fig.~\ref{fig:splitting}(c), thin green line] and the analytical result from a perturbation theory calculation to first order in $\Omega$ [Fig.~\ref{fig:splitting}(c), thick red line], which are discussed in the following. All system parameters which are relevant to the calculations were extracted from separate experiments. The drive amplitude seen by the transmon qubit was calibrated by fitting the $\Omega$ dependence of $\Delta_{f0g1}$. The calibration routine for the coherent drive strength is further discussed in Section~$\mathrm{\ref{sec:ACStark}}$.
\section{Calculating the coupling strength $\tilde{g}$ }
\label{sec:method}

\begin{figure}[ht!]
  \centering
  \includegraphics[width=85mm,trim=0mm 55mm 0mm 20mm, clip]{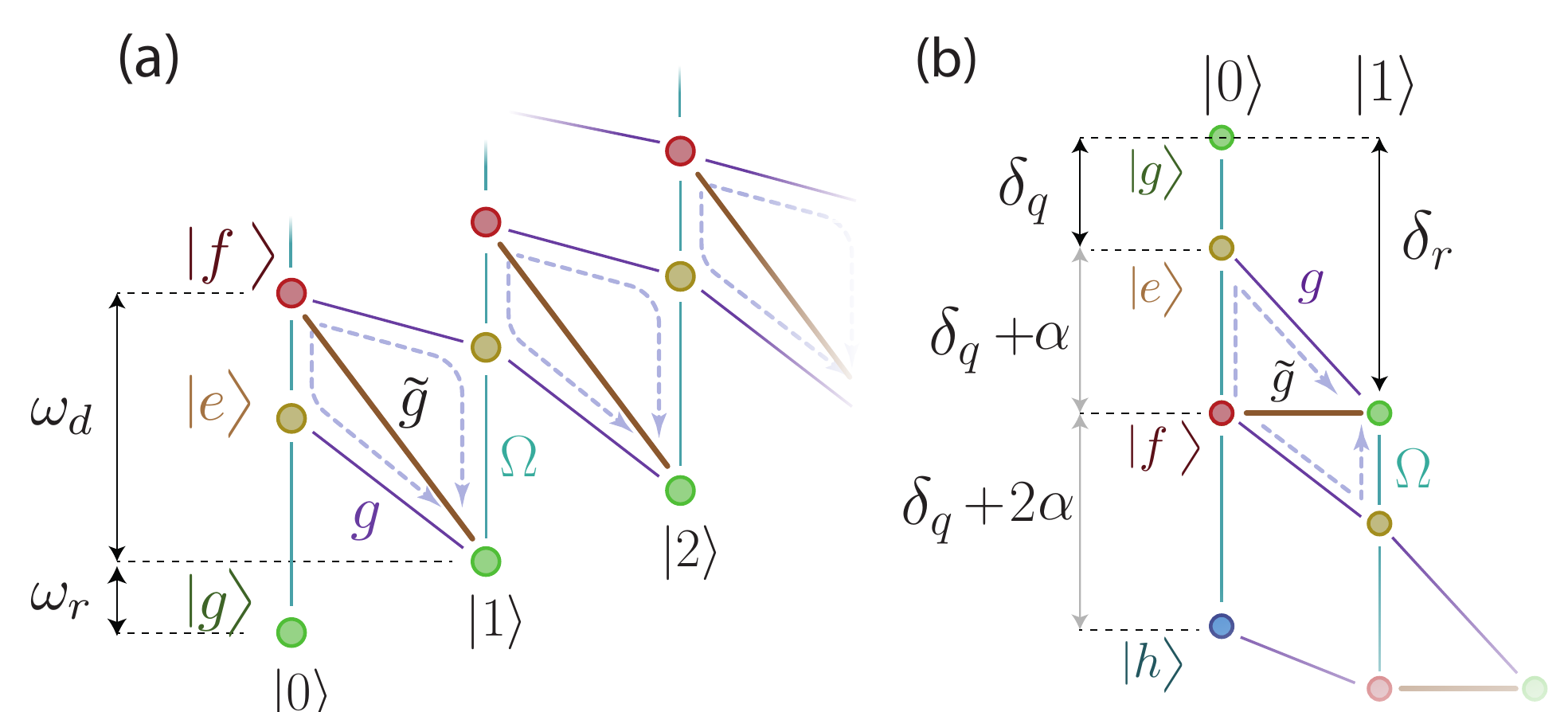}
  \caption{Schematic energy level diagram of the system (a) in the laboratory frame and
   (b) in the rotating frame of the drive frequency $\omega_{d}$ that compensates for the energy difference between $\ket{f0}$ and $\ket{g1}$. The effective tunable coupling $\tilde{g}$ between the bare states $\ket{f0}$ and $\ket{g1}$ can be understood as a second order cavity-assisted Raman process, indicated by the dashed arrows.}
  \label{fig:energylevels}
\end{figure}

In a reference frame rotating at the frequency $\omega_{d}$, the Hamiltonian for the transmon coupled to a resonator mode can be written as a sum of an N-level Jaynes-Cummings term and a coherent drive term, $H=H_{\mathrm{JC}}+H_{d}$. Here \cite{Koch2007},
\begin{eqnarray}
  \label{eq:hamiltonianramanRF}
  &H_{\mathrm{JC}} =& \delta_r a^\dagger a +\delta_q b^\dagger b + \frac{\alpha}{2} b^\dagger b^\dagger b b \\
  \nonumber &&+ g(a b^\dagger + a^\dagger b),\\
  \nonumber \mathrm{and}&&\\
  &H_{d}=& \frac{\Omega(t)}{2}\left(\mathrm{e}^{\mathrm{i} \phi} b + h.c.\right)\, ,\label{eq:hamiltonianramanRFdrive}
  \end{eqnarray}
where $\delta_r \equiv \omega_r^{0} - \omega_d$ and $\delta_{q} \equiv \omega_{ge}^{0}-\omega_d$ as the resonator-drive and transmon-drive detunings, respectively. $\omega_r^{0}$ and $\omega_{ge}^{0}$ denote the bare quantities associated with $\omega_r$ and $\omega_{ge}$ defined above (Fig.~$\ref{fig:energylevels}$). The operator $a\,(a^{\dagger})$ is the annihilation (creation) operator for the resonator mode and $b\,(b^{\dagger})$ its analogue for the transmon $b \equiv |g\rangle\langle e| + \sqrt{2} |e\rangle\langle f|+\sqrt{3} |f\rangle\langle h|+\ldots\,$ when treated as an anharmonic oscillator at frequency $\omega_{ge}^{0}$ \cite{Koch2007}. The Jaynes-Cummings type interaction \cite{Blais2004} couples the states which have the same total number of excitations, while the coherent drive field with time-dependent amplitude $\Omega(t)$, frequency $\omega_d$, and phase $\phi$ couples neighboring pairs of transmon states. In the following, we omit the time dependence of $\Omega(t)$ for notational clarity. The above Hamiltonian is valid in the transmon limit where $E_\mathrm{J}/E_\mathrm{C}\gg 1$.

We consider the system in the dispersive regime, that is, $\Delta = \omega_{ge}-\omega_r\gg g$. To calculate the tunable effective coupling strength, we rewrite the Hamiltonian $H$ in the eigenbasis of $H_{\mathrm{JC}}$, and treat the coherent drive term $H_{d}$ perturbatively, expanding the solution in powers of the small parameter $\Omega$. Note that $\ket{i,j}_D$ ($i=g,e,f,h,\dots$ and $j=0,1,2,\dots$) is used for the eigenstates of $H_{\mathrm{JC}}$, and $\ket{i,j}$ for the bare qubit-resonator states. The drive Hamiltonian $H_d$ can be written in the $\ket{i,j}_D$ basis as
\begin{equation}
\nonumber H_{d}=\sum_{ij,kl} \frac{\Omega}{2} \ket{i,j}_{D}\bra{k,l}_{D} \left\{\bra{i,j}_{D}(\mathrm{e}^{\mathrm{i}\phi} b+\mathrm{e}^{-\mathrm{i}\phi} b^{\dagger})\ket{k,l}_{D}\right\},
\label{eq:hamiltonianramanJC}
\end{equation}

When the resonance condition in Eq.~(\ref{eq:resonantconditionstark}) is satisfied the main contribution to the evolution of the system comes from the terms coupling the resonant states $\ket{g,l+1}_D$ and $\ket{f,l}_D$ while the terms describing off-resonant transitions in Eq.~($\ref{eq:hamiltonianramanJC}$) can be neglected. With this rotating-wave-type approximation, the drive Hamiltonian becomes
\begin{eqnarray}
\nonumber H_{d}&\approx& \sum_{l} \frac{\Omega}{2} \ket{g,l+1}_{D}\bra{f,l}_{D}\\ && \left\{\bra{g,l+1}_{D}(\mathrm{e}^{\mathrm{i}\phi} b\right.
\left.+\mathrm{e}^{-\mathrm{i}\phi} b^{\dagger})\ket{f,l}_{D}\right\}.
\label{eq:RedHam}
\end{eqnarray}

To show that the effective coupling between the transmon and the resonator is indeed of the Jaynes-Cummings type, we note that the dressed states $\ket{f,l}_{D}$ and $\ket{g,l+1}_{D}$ are given, up to first order in $g$, by
\begin{eqnarray}
\nonumber\ket{f,l}_{D}&=&|f,l\rangle-\frac{g\sqrt{2(l+1)}}{\Delta+\alpha}|e,l+1\rangle\\
&&+\frac{g\sqrt{3l}}{\Delta+2\alpha}|h,l-1\rangle\label{eq:JCeigenstates0},\\
\ket{g,l+1}_{D}&=&|g,l+1\rangle+\frac{g\sqrt{l+1}}{\Delta}|e,l\rangle,
\label{eq:JCeigenstates}
\end{eqnarray}
where the coupling $g(ab^\dagger~+~a^\dagger b)$ is considered as a perturbation to the uncoupled transmon-resonator system (i.e.,~$\left(g/\Delta\right)^{2}~\ll~1$). Using this approximation for the dressed states, we calculate the matrix element

\begin{equation}
\tilde{g}_{l}\equiv \bra{g,l+1}_{D}\,H_{d}\ket{f,l}_{D} \approx g\Omega \mathrm{e}^{\mathrm{i} \phi} \sqrt{\frac{l+1}{2}}\frac{\alpha}{\Delta(\Delta+\alpha)}\, ,
 \label{eq:effectiveg}
\end{equation}
which represents the coupling strength between dressed states $\ket{f,l}_{D}$ and $\ket{g,l+1}_{D}$. As expected, $\tilde{g}_{l}$ is tunable both in phase and amplitude because of its dependence on the complex drive strength $\Omega e^{i\phi}$.

Next, defining the qubit raising and lowering operators
\begin{equation}
\tilde{\sigma}\equiv \sum_{l}\ket{g,l}_D\bra{f,l}_D\,\qquad \tilde{\sigma}^{\dagger}\equiv \sum_{l}\ket{f,l}_D\bra{g,l}_D,
\label{eq:qubitraise}
\end{equation}
and the dressed photon annihilation operator
\begin{equation}
\tilde{a}~\equiv~\sum_{l,j}\sqrt{l+1}~\ket{j,l}_{D}\bra{j,l+1}_{D},
\end{equation}
the drive Hamiltonian in Eq.~($\ref{eq:RedHam}$) can be written in the Jaynes-Cummings form
\begin{eqnarray}
\tilde{H}_{d}&\approx&\tilde{g}\tilde{a}^{\dagger}\tilde{\sigma_l}+\tilde{g}^{*}\tilde{a}\tilde{\sigma_l}^{\dagger}
\label{eq:JCHAM}
\end{eqnarray}
with $\tilde{g}~\equiv~\tilde{g}_{0}~\approx~g\Omega \mathrm{e}^{\mathrm{i}\phi}\alpha/\left(\sqrt{2}\Delta(\Delta+\alpha)\right)$. The absolute value of the coupling $\tilde{g}$ in Eq. ($\ref{eq:JCHAM}$) describes the splitting observed in Fig.~$\ref{fig:splitting}$(c).

It is instructive to compare the tunable transmon-resonator coupling $\tilde{g}$ to the effective coupling $g_{\Lambda}=g\Omega \mathrm{e}^{\mathrm{i} \phi}/\Delta$ between the two degenerate states of a $\Lambda$ system obtained from the adiabatic elimination technique \cite{Gerry1990,Alexanian1995,Wu1997,Brion2007}. Most notably, the coupling strength for the transmon-resonator system is lower than that of the $\Lambda$ system, $\tilde{g}<g_{\Lambda}$. This result follows from the opposite signs of perturbative contributions from $\ket{e,l}$ and $\ket{e,l+1}$ in Eq.~($\ref{eq:JCeigenstates0}$-$\ref{eq:JCeigenstates}$). Physically, this effect arises from the destructive interference of the two second order transition paths that couple degenerate levels $\ket{f0}_{D}$ and $\ket{g1}_{D}$ (Fig.~$\ref{fig:energylevels}$). Indeed, the only reason that the coupling $\tilde{g}$ does not vanish completely is the anharmonicity $\alpha$ of the transmon qubit, which results in a difference in the magnitude of perturbative contributions from $\ket{e,l}$ and $\ket{e,l+1}$. On the other hand, in a $\Lambda$ system, there is only one transition path coupling the degenerate levels, and consequently there are no interference effects. The correspondence between the $\Lambda$ and the transmon systems is easy to see in the limit of large anharmonicity, when $\tilde{g}(\alpha\rightarrow \infty)=g_{\Lambda}$.

As shown in Fig.~$\ref{fig:splitting}$(c), the first order perturbation theory gives a satisfactory approximation to the tunable transmon-resonator coupling for small drive amplitudes which satisfy $\left(\Omega/2(\Delta+\alpha)\right)^{2}\ll 1$. However, the approximation to $\tilde{g}$ breaks down when this inequality is no longer satisfied, and the first order approximations to $\ket{f0}_D$ and $\ket{g1}_D$ in Eq.~($\ref{eq:JCeigenstates0}$-$\ref{eq:JCeigenstates}$) lose their validity. In particular, we observe in the experiment that the first order approximation starts to break down as $\Omega/2\pi$ is increased above approximately 0.2~GHz for $\Delta/2\pi=$~0.979~GHz, i.e., $\left(\Omega/2(\Delta+\alpha))\right)^{2}\approx 0.025$.

\section{AC Stark Shift and Drive Power Calibration}
\label{sec:ACStark}

To determine the conversion factor between the applied drive power and the drive amplitude $\Omega$ seen by the transmon qubit we fit the observed Stark shift to a perturbative expression for $\Delta_{f0g1}$, where all the parameters other than $\Omega$ can be determined from separate measurements. In the following, we discuss the resolvent method used to obtain such a perturbative expression for $\Delta_{f0g1}$.

The resolvent method allows for a systematic approximation of $\Delta_{f0g1}$ for increasing orders of interaction in both $g$ and $\Omega$. The Hamiltonian $H = H_0 + H_I$ consists of a bare part $H_0$ and an interaction part $H_I$ and its resolvent is defined as \cite{kaku1993QFT}:
\begin{equation}
G(z)=\frac{1}{z-H},
\end{equation}
where $z$ is a complex variable. In particular, the eigenenergies of $H$ are given by the poles of $G(z)$. Specifically for our transmon-resonator system, we split the driven Jaynes-Cummings Hamiltonian given by Eqs.~(\ref{eq:hamiltonianramanRF}) and (\ref{eq:hamiltonianramanRFdrive}) into
\begin{eqnarray}
&H_0=&\delta_r a^\dagger a +\delta_q b^\dagger b + \frac{\alpha}{2} b^\dagger b^\dagger b b\\
\nonumber \mathrm{and}&&\\
&H_I&=g\,a b^\dagger+\frac{\Omega(t)}{2} \mathrm{e}^{\mathrm{i} \phi} b + h.c.\, ,
\end{eqnarray}
where $g$ and $\Omega$ are real numbers. With this notation, we can rewrite the resolvent as
\begin{equation}
G(z)=\frac{1}{z-H_0}\sum_{n=0}^\infty \left(\frac{H_I}{z-H_{0}}\right)^n.
\label{eq:resolventpert}
\end{equation}
If the states $\ket{\phi_i}$ and $\ket{\phi_j}$ are two degenerate eigenvectors of the bare Hamiltonian satisfying $H_0\ket{\phi_{i,j}}=E\ket{\phi_{i,j}}$, the resolvent operator restricted to the two-dimensional space spanned by them can be written in the final form
\begin{equation}
G(z)=\frac{1}{z-H_0-\Sigma(z)},
\label{eq:resolventfinal}
\end{equation}
where the operators $H_0$ and $\Sigma(z)$ are to be understood as acting on the two-dimensional space and the matrix elements $\Sigma_{kl}(z)=\langle\phi_k|\Sigma(z)|\phi_l\rangle$ for $k,l\in\{i,j\}$ are given by
\begin{widetext}
\begin{eqnarray}
\nonumber\Sigma_{kl}(z)=&&\bra{\phi_k}H_I\ket{\phi_l}+\sum_{m\neq i,j}\bra{\phi_k}H_I\ket{\phi_m}\frac{1}{z-E_m}\bra{\phi_m}H_I\ket{\phi_l}\\
&&+\sum_{m,m'\neq i,j}\bra{\phi_k}H_I\ket{\phi_m}\frac{1}{z-E_m}\bra{\phi_m}H_I\ket{\phi_{m'}}\frac{1}{z-E_m'}\bra{\phi_{m'}}H_I\ket{\phi_l}
+\cdots\, ,
\label{eq:pathsum}
\end{eqnarray}
\end{widetext}
which is the weighted sum of all transition paths coupling states $\ket{\phi_k}$ and $\ket{\phi_l}$ through intermediate bare eigenstates $\{\ket{\phi_m}\}_{m=1}^{\infty}$ excluding $\ket{\phi_i}$ and $\ket{\phi_j}$. Equation~($\ref{eq:resolventfinal}$) is obtained by inserting an appropriate number of copies of the identity $\mathbb{I}=\sum_{m}\ket{\phi_m}\bra{\phi_m}$ into each term of the sum in Eq.~($\ref{eq:resolventpert}$), and by noticing that the identity
\begin{eqnarray}
\nonumber&&\bra{\phi_k}\left(\frac{1}{z-H_0}\sum_{n=0}^\infty \left(\frac{H_I}{z-H_{0}}\right)^n\right)\ket{\phi_l}\\
&&=\bra{\phi_k}\left(\frac{1}{z-H_0}\sum_{n=0}^\infty \left(\frac{\Sigma}{z-H_{0}}\right)^n\right)\ket{\phi_l}
\label{eq:identity}
\end{eqnarray} holds if $k,l\in\{i,j\}$. 

\begin{figure}[hb!]
  \centering
  \includegraphics[width=86mm,trim=0mm 0mm 0mm 0mm, clip]{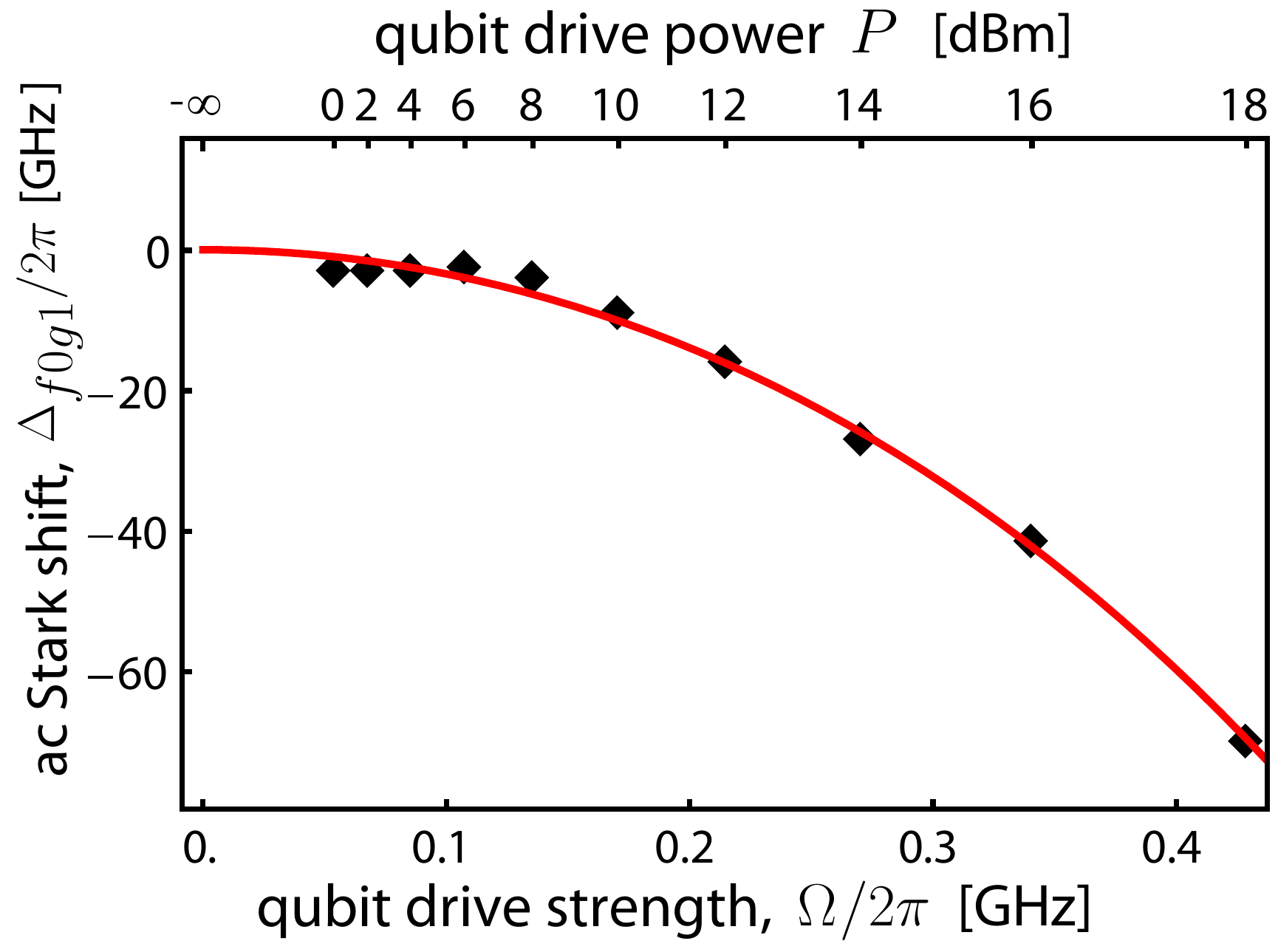}
  \caption{Measured values of $\Delta_{f0g1}$ for increasing drive amplitudes $\Omega$. A fit to the expression in Eq.~($\ref{eq:ACSystem}$) obtained from the resolvent method (solid line) where the perturbative series is truncated at the the sixth order in $\Omega$. The fit gives the conversion factor between the applied drive power and the applied drive amplitude seen by the transmon. The scale on the top indicates the drive power at the signal generator where each tick stands for the drive power at which the data point is obtained.}
  \label{fig:ACStark}
\end{figure}

Eq.~(\ref{eq:resolventfinal}) shows that $H_0+\Sigma(z)$ can be interpreted as an effective Hamiltonian $H^{\mathrm{eff}}(z)$ describing the evolution of the system in the two-dimensional subspace spanned by $\ket{\phi_i}$ and $\ket{\phi_j}$.
The resonance condition ($\ref{eq:resonantconditionstark}$), which corresponds to the avoided crossing in Fig. $\ref{fig:splitting}$, can now be expressed as 
\[
 z - H^{\mathrm{eff}}_{g1g1}(z) = z - H^{\mathrm{eff}}_{f0f0}(z) = 0,
\]
which is to be satisfied when the bare energies $E_{f0}$ and $E_{g1}$ are separated by a detuning $\Delta_{f0g1}(\Omega)+\Delta_{\mathrm{JC}}$. Here the constant term $\Delta_{\mathrm{JC}}$ represents the $\Omega$-independent renormalization of the bare energies due to the Jaynes-Cummings coupling $g$ only, such that $\Delta_{f0g1}(0)=0$. Thus, we can calculate the detuning $\Delta_{f0g1}(\Omega)$ by iteratively solving the coupled equations
\begin{eqnarray}
\nonumber && z-E_{g1}-\Sigma_{g1g1}(z,\Omega)=0\\
\nonumber && z-E_{g1}+\Delta_{f0g1}(\Omega)+\Delta_{\mathrm{JC}}-\Sigma_{f0f0}(z,\Omega)=0,\\
\label{eq:ACSystem}
\end{eqnarray}
with the initial value $z=0$. Fitting the resulting expression for $\Delta_{f0g1}(\Omega)$ to the experimental data (see~Fig.~$\ref{fig:ACStark}$) provides the conversion factor $k$ between the applied drive power and the drive amplitude $\Omega$ seen by the transmon. This conversion factor is also used to compute the qubit drive strength value in Fig.~$\ref{fig:splitting}$(c), giving good agreement between our measurement and perturbation theory calculation.

\section{Numerical and Analytical results on Time-Dependent Vacuum
  Rabi oscillations}
\label{sec:numerical}

The effective transmon-resonator coupling leads to a coherent exchange of excitations (swap) between the states $\ket{f,l}_{D}$ and $\ket{g,l+1}_{D}$. Achieving a high fidelity for this swap operation is crucial for further applications of the microwave-induced transmon-resonator coupling in the context of quantum computing. In this section, we first derive an analytical expression for the fidelity of the swap operation when the external drive is turned on and off instantaneously. Then we use numerical simulations to account for the effects arising from pulse profiles that vary slowly with respect to $\Delta$. We conclude that the swap operation can be realized with very high fidelity for realistic pulse profiles.

When the drive is turned on and off instantaneously, one can derive an analytical expression for the Rabi oscillations in the population in $\ket{g1}_D$, given the initial state $\ket{f0}_D$. The population in $\ket{g1}_{D}$ at time $t$, denoted $P_{g1D}(t)$, is given by the modulus square of the overlap between the time evolved initial state $\ket{f0(t)}_D$ and the target state $\ket{g1}_{D}$
\begin{equation}
P_{g1D}(t)=\left|\bra{g1}_{D}\,\ket{f0(t)}_{D}\right|^{2},
\end{equation}
where $\ket{f0(t)}_{D}~=~U(t)\ket{f0}_{D}$ with the unitary time evolution operator $U(t)\equiv e^{-i H t}$.

To express $P_{g1D}(t)$ in a convenient form, we expand the initial and target state in the eigenbasis of the full Hamiltonian
\begin{eqnarray}
\ket{f0}_D=\sum_n \alpha_n \ket{\Phi_n}\\
\ket{g1}_D=\sum_n \beta_n \ket{\Phi_n},
\label{eq:expansion}
\end{eqnarray}
where the eigenstates $\ket{\Phi_n}$ satisfy $H\ket{\Phi_n}=\epsilon_n\ket{\Phi_n}$, and the coefficients $\alpha_n$ and $\beta_n$ are defined as $\langle \Phi_n\ket{f0}_D$ and $\langle \Phi_n\ket{g1}_D$, respectively. In this new basis, the time-dependent population in $\ket{g1}_D$ is
\begin{eqnarray}
P_{g1D}(t)&=&\sum_{n}\alpha_{n}^{2}\beta_n^2\label{eq:fidelity1}\\
\nonumber&&+2\sum_{n<m}|\alpha_{m}^{*}\beta_{m}\alpha_{n}^{*}\beta_{n}|\cos\left[(\epsilon_{n}-\epsilon_{m})t+\theta_{nm}\right],
\end{eqnarray}
\noindent
where $\theta_{nm}$ is the phase of ($\alpha_{m}^{*}\beta_{m}\alpha_{n}^{*}\beta_{n}$).

Equation~($\ref{eq:fidelity1}$) is an exact and useful relation between the solution to the full Hamiltonian (which can be approximated using perturbation theory) and the rate and fidelity of the population exchange between $\ket{f0}_{D}$ and $\ket{g1}_{D}$. In particular, the maximum fidelity $\mathcal{F}$ of the swap operation between $\ket{f0}_{D}$ and $\ket{g1}_{D}$ can be extracted from Eq.~($\ref{eq:fidelity1}$) as $\mathrm{\cal{F}} = 4 |\alpha_{+}\alpha_{-}\beta_{+}\beta_{-}|$ and its rate as $2\tilde{g} = \epsilon_{+}-\epsilon_{-}$, where the subscripts $\pm$ stand for the two eigenstates of $H$ which have the largest overlaps with $\ket{f0}_D$ and $\ket{g1}_D$. We denote these eigenstates with $\ket{\Phi_{\pm}}$. When the resonance condition in Eq.~($\ref{eq:resonantconditionstark}$) is satisfied and there are no other states whose energies are close to that of $\ket{f0}_{D}$ or $\ket{g1}_{D}$, i.e., $\left(\Omega/2/(\Delta+\alpha)\right)^{2}\ll~1$, these eigenstates are simply the two polariton states $\ket{\Phi_{\pm}}=1/\sqrt{2}\left(\ket{f0}_{D}\pm\ket{g1}_{D}\right)$. In this weak drive limit, Eq.~($\ref{eq:fidelity1}$) implies that the population exchange between $\ket{f0}_D$ and $\ket{g1}_D$ occurs with unit fidelity $\mathcal{F}=1$.

\begin{figure}[h!]
\centering
\includegraphics[width=85mm]{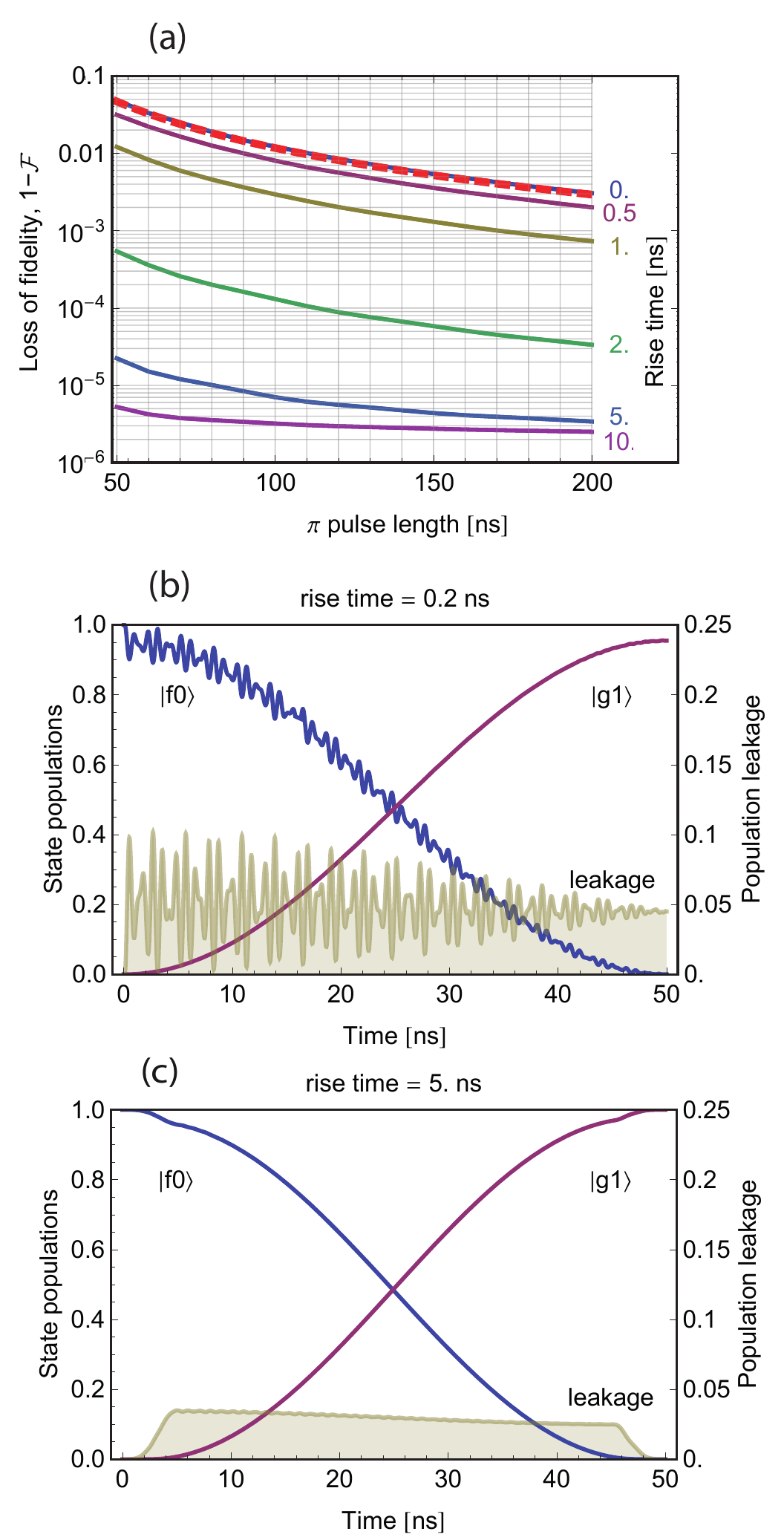}
\caption{(a) Dependence of the loss of fidelity $1-\cal{F}$ on the length of the $\pi$ pulse and its rise time. The transmon-resonator detuning is set to $\Delta/2\pi=0.979$ GHz, the anharmonicity is $\alpha/2\pi~=-~0.376$ GHz, and $g/2\pi~=~65$~MHz. The solid lines show the results of the numerical simulations. The theoretical prediction of Eq.~($\ref{STIFID}$) is shown in dashed red. (b)-(c) Time evolution of the populations $p_{f0}$, $p_{g1}$ in $\ket{f0}_D$, $\ket{g1}_D$, and the leakage to other levels given by $1-p_{g1}-p_{f0}$ for a $50\,\mathrm{ns}$ drive pulse with rise times of (b) $0.2\,\mathrm{ns}$ and (c) $5.0\,\mathrm{ns}$.}
\label{VISFIT}
\end{figure}

For realistic amplitudes and detunings a reduction in $\cal{F}$ is caused by the population leakage out of the initial and target states. To calculate the $\Omega$-dependence of the fidelity, we use the second order corrections to the eigenstates $\ket{\Phi_{\pm}}$ induced by the perturbation $\tilde{H}_{d}$ (see appendix \ref{app:Nonad}). As a result, the fidelity of the swap operation between $\ket{f0}_{D}$ and $\ket{g1}_{D}$ is
\begin{equation}
\mathrm{\cal{F}}= 1-\left(\left(\frac{\Omega/2}{\Delta+\alpha}\right)^{2}+\left(\frac{\sqrt{2}\Omega/2}{\Delta}\right)^{2}+\left(\frac{\sqrt{3}\Omega/2}{\Delta-\alpha}\right)^{2}\right),
\label{STIFID}
\end{equation}
up to second order in $\Omega/\Delta$. Notice that the reduction in fidelity is caused by population leakage out of the states $\ket{f0}_D$ and $\ket{g1}_D$ to neighboring states. 
We observe that the simulated fidelity shown in Fig.~\ref{VISFIT}a for a drive pulse with zero rise time is in good agreement with the analytical calculation in Eq.~(\ref{STIFID}). The simulation of the transmon dynamics in the absence of decoherence was performed by solving the Schr\"{o}dinger equation with the Hamiltonian given by Eqs.~(\ref{eq:hamiltonianramanRF}) and (\ref{eq:hamiltonianramanRFdrive}). The underlying Hilbert space was truncated to four energy levels of the transmon and four Fock states of the resonator.

The calculation above is based on the assumption that the drive pulse is turned on an off instantaneously and the obtained fidelity is therefore valid in the limit of an ideal square pulse. If, however, the pulse is ramped up and down gradually, the process becomes adiabatic with respect to the off-resonant transitions. Simulation results presented in Fig.~\ref{VISFIT} show that this effect improves the fidelity significantly.

The shape of the drive pulse $\Omega(t)$ for the simulations is chosen such that the effective coupling $\tilde{g}(t)$ has the form
\begin{align*}
  \tilde{g}(t) &= \tilde{g}_{\mathrm{max}} &\text{ for }\Delta t < t < T - \Delta t,\\
  &= \tilde{g}_{\mathrm{max}}\sin^2(\pi t/2\Delta t) &\text{ for }t\le\Delta t,\\
  &= \tilde{g}_{\mathrm{max}}\sin^2(\pi (T-t)/2\Delta t) &\text{ for }t\ge T-\Delta t,
\end{align*}
where $T$ denotes the length of the pulse, $\Delta t$ the rise time and the amplitude $\tilde{g}_{\mathrm{max}}$ is chosen such that $\int_{0}^{T}\tilde{g}(t)\,\mathrm{d}t = \pi/2$, resulting in a $\pi$-flip between the states $|f0\rangle_D$ and $|g1\rangle_D$. The frequency of the pulse is varied in time to exactly cancel the variations in the amplitude-dependent ac Stark shift.

To generate the correct drive pulse for the simulation, we needed to calculate the ac Stark shift and the effective coupling $\tilde{g}$ with good accuracy. For this reason, we decided to use a numerical procedure based on diagonalization of the Hamiltonian which is described in more detail in Appendix \ref{app:numStark}.
While this method is more accurate than the analytical expression obtained using perturbation theory, the latter is a more time efficient way of determining the behaviour of ac Stark shift as a function of other system parameters (i.e. $\Delta$, $\alpha$, and $g$).

With pulses generated by this method, the simulations show that the population swap between $|f0\rangle_D$ and $|g1\rangle_D$ can be realized with essentially unit fidelity ($>0.99999$) when decoherence is neglected. In current state-of-the-art experiments, the achievable fidelities will therefore be limited mainly by coherence times of the transmons or imperfections in the generated drive pulses.

\section{Conclusion}

In conclusion, we have experimentally demonstrated a cavity-assisted Raman process to realize a tunable transmon-resonator coupling in a superconducting circuit QED architecture. This effective coupling is induced by coherently driving a transmon qubit. Its maximal value is only determined by the maximum qubit drive power that can be applied to the qubit. The measured data is in very good agreement with a perturbative calculation of the coupling strength $\tilde{g}$ between $\ket{f0}_D$ and $\ket{g1}_D$. Our calculations show that the strength of the Raman transition is reduced in comparison to the one implemented in a $\Lambda$ level system \cite{Gerry1990,Law1997} by the destructive interference between the two second order transition paths which couple $|f0\rangle_{D}$ and $|g1\rangle_{D}$.

We also determined the fidelity of the swap operation which utilized the cavity-assisted Raman process, by both numerical and analytical means. A second order degenerate perturbation calculation shows that the fidelity of the population exchange strongly depends on the population leakage from $\ket{f0}_D$ and $\ket{g1}_D$ to the closest lying states [see Eq.~($\ref{STIFID}$)], a result which is also supported by numerical simulations (Fig.~$\ref{VISFIT}$). We found that the fidelity of the swap operation is expected to be very close to unity in the absence of decoherence effect and in realistic systems will most likely be limited by the coherence time of the transmon.

We would like to thank Alexandre Blais for useful comments on the manuscript. This work was supported by the Swiss National Science Foundation (SNF), Project 150046, by the National Center
of Competence in Research ``Quantum Science and Technology'' and by ETH Zurich.

\appendix

\section{Numerical calculation of the ac Stark shift}
\label{app:numStark}
In this appendix, we describe the evolution of the transmon under the approximation of slowly varying drive pulses. We then define the ac Stark shift operationally as the shift of the drive frequency required to implement a perfect swap between the states $|f0\rangle_D$ and $|g1\rangle_D$. Finally, we describe a method for calculating the ac Stark shift numerically.

If the drive amplitude $\Omega(t)$ is varied slowly compared to the energy separation of $|f0\rangle_D$ and $|g1\rangle_D$ from other energy eigenstates, the system evolves adiabatically with respect to the off-resonant transitions. Therefore, if the system is initially prepared in a superposition of $|f0\rangle_D$ and $|g1\rangle_D$, it remains at all times in the subspace $\mathcal{S}(\Omega)$ spanned by the two instantaneous eigenstates $|\Phi_1(\Omega)\rangle$ and $|\Phi_2(\Omega)\rangle$ corresponding to $|f0\rangle_D$ and $|g1\rangle_D$. 

To describe the evolution of the state vector $|\Psi(t)\rangle$ in $\mathcal{S}(\Omega)$, we first introduce a mapping which connects the subspaces $\mathcal{S}(\Omega)$ for different values of $\Omega$. Using this transformation, the action of the full Hamiltonian for any $\Omega$ is mapped to an effective Hamiltonian acting on $\mathcal{S}(0)$, that is, the subspace spanned by $|f0\rangle_D$ and $|g1\rangle_D$. Using this formalism, we calculate the Stark shift as the shift of the drive frequency needed to keep the effective Hamiltonian in the ``resonant form'' $\tilde{g}\sigma_x \equiv \tilde{g}(|f0\rangle_D\langle g1|_D+\mathrm{H.c.})$. If the drive frequency is not adjusted, the effective Hamiltonian contains a term proportional to $\sigma_z$ which prevents us from realizing a perfect swap operation between the two states.

We start by introducing a linear map $M_\Omega: \mathcal{S}(0) \to \mathcal{S}(\Omega)$ defined as
\begin{equation}\label{eq:MOpDef}
  M_\Omega = \lim_{\Delta \Omega\to 0} P(\Omega)P(\Omega-\Delta\Omega)\ldots P(2\Delta\Omega)P(\Delta\Omega)P(0),
\end{equation}
where $P(\Omega) = \sum_{i=1,2}{|\Phi_i(\Omega)\rangle\langle\Phi_i(\Omega)|}$ is a projector onto $\mathcal{S}(\Omega)$. This map represents a continuous series of projections onto the subspaces $\mathcal{S}(x)$ for $x$ varying from 0 to $\Omega$. 

As the state vector $|\Psi(t)\rangle$ evolves adiabatically and therefore lies in $\mathcal{S}(\Omega(t))$, we can write it in the form
\[
  |\Psi(t)\rangle = M_{\Omega(t)} |\psi(t)\rangle,
\]
where $|\psi(t)\rangle$ is some vector evolving in the subspace $\mathcal{S}(0)$ spanned by $|f0\rangle_D$ and $|g1\rangle_D$. In this way, we have reduced the problem of finding the evolution of the state vector $|\Psi(t)\rangle$ in a changing subspace $\mathcal{S}(\Omega(t))$ to that of finding the evolution of the directly related vector $|\psi(t)\rangle$ in a fixed subspace $\mathcal{S}(0)$.

Since $|\Psi(t+\mathrm{d}t)\rangle$ lies in $\mathcal{S}(\Omega(t+\mathrm{d}t))$, it can be expressed as
\begin{align*}
  |\Psi(t+\mathrm{d}t)\rangle =& P(\Omega(t+\mathrm{d}t)) |\Psi(t+\mathrm{d}t)\rangle\\
  =& P(\Omega(t+\mathrm{d}t))\exp(-\mathrm{i}H(t)\mathrm{d}t)|\Psi(t)\rangle,
\end{align*}
resulting in the following evolution equation for $|\psi(t)\rangle$:
\begin{align}
  M_{\Omega(t+\mathrm{d}t)} &|\psi(t+\mathrm{d}t)\rangle = \label{eq:towardsEffHam}\\
  &P(\Omega(t+\mathrm{d}t))\exp(-\mathrm{i}H(t)\mathrm{d}t)
  M_{\Omega(t)} |\psi(t)\rangle\nonumber
\end{align}
Under the reasonable assumption that the subspace $\mathcal{S}(\Omega)$ changes smoothly with $\Omega$, it can be shown that $M_\Omega$ preserves vector norms in $\mathcal{S}(\Omega)$. It follows that $M_\Omega^{\dagger}$, which is an infinite product of projectors analogous to $M_\Omega$ but in the reverse order, is the inverse of $M_\Omega$. Hence, after dropping the projector $P(\Omega(t+\mathrm{d}t))$ from the right-hand side of Eq.~(\ref{eq:towardsEffHam}) as well as from the product form of $M_{\Omega(t+\mathrm{d}t)}$ (cf.~Eq.~(\ref{eq:MOpDef})) on its left-hand side, we multiply the equation by $M_{\Omega(t)}^{\dagger}$ to obtain
\[
  |\psi(t+\mathrm{d}t)\rangle = 
  M_{\Omega(t)}^{\dagger} \exp(-\mathrm{i}H(t)\mathrm{d}t)
  M_{\Omega(t)}^{\vphantom{\dagger}} |\psi(t)\rangle
\]
which we transform into the differential form
\[
  \frac{\mathrm{d}}{\mathrm{d}t} |\psi(t)\rangle = 
  -\mathrm{i} M_{\Omega(t)}^{\dagger} H(t) M_{\Omega(t)}^{\vphantom{\dagger}} |\psi(t)\rangle.
\]

The evolution of the vector $|\psi(t)\rangle$ is therefore governed by an effective Hamiltonian
\begin{equation}\label{eq:effHam}
  H_{\mathrm{eff}}(t) = M_{\Omega(t)}^{\dagger} H(t) M_{\Omega(t)}^{\vphantom{\dagger}}
\end{equation}
acting on $\mathcal{S}(0)$, resulting in the evolution operator $U(t_{\mathrm{f}},t_{\mathrm{i}}) = \mathcal{T}\exp\int_{t_{\mathrm{i}}}^{t_{\mathrm{f}}}{(-\mathrm{i}H_{\mathrm{eff}}(t))\,\mathrm{d}t}$. If we assume that the drive amplitude at the initial time $t_\mathrm{i}$ and the final time $t_\mathrm{f}$ is zero, we have $M_{\Omega(t_\mathrm{i})} = M_{\Omega(t_\mathrm{f})} = P(0)$ and hence $|\psi(t_{\mathrm{i}})\rangle = |\Psi(t_{\mathrm{i}})\rangle$ and $|\psi(t_{\mathrm{f}})\rangle = |\Psi(t_{\mathrm{f}})\rangle$. Then we can directly write down the evolution from $|\Psi(t_{\mathrm{i}})\rangle$ to $|\Psi(t_{\mathrm{f}})\rangle$:
\[
  |\Psi(t_{\mathrm{f}})\rangle = 
  U(t_{\mathrm{f}},t_{\mathrm{i}}) |\Psi(t_{\mathrm{i}})\rangle.
\]

This result allows to determine the ac Stark shift as the amplitude-dependent shift of the drive frequency needed to realize a perfect swap operation between $|f0\rangle_D$ and $|g1\rangle_D$. For this, the effective Hamiltonian has to have the form
\begin{equation}\label{eq:heffForm}
  H_{\mathrm{eff}}(t) = E_{\mathrm{offset}}(t)\mathbbm{1} + \tilde{g}(t)
  (|f0\rangle_D\langle g1|_D+\mathrm{H.c.}),
\end{equation}
where the overall energy shift $E_{\mathrm{offset}}(t)$ leading only to an overall phase shift is omitted since it is physically irrelevant. This equation is equivalent to the requirement that the equal superposition states $|\varphi_{1,2}\rangle = (|f0\rangle_D \pm |g1\rangle_D)/\sqrt{2}$ are eigenstates of $H_{\mathrm{eff}}(t)$ and therefore, by virtue of Eq.~(\ref{eq:effHam}), that $M_{\Omega(t)}|\varphi_{1,2}\rangle$ are eigenstates of $H(t)$ which we have previously denoted by $|\Phi_{1,2}(\Omega)\rangle$. In other words,
\begin{equation}\label{eq:eigenvecParTransport}
  |\Phi_{i}(\Omega)\rangle = M_{\Omega}|\varphi_{i}\rangle.
\end{equation}

This equation can be solved for $M_{\Omega}$. However, since our goal is to determine the ac Stark shift, we need an equation for the Hamiltonian instead. To get it, we transform Eq.~(\ref{eq:eigenvecParTransport}) into a differential form. By substituting $\Omega\to \Omega+\mathrm{d}\Omega$, we find the following relation between $|\Phi_{i}(\Omega+\mathrm{d}\Omega)\rangle$ and $ |\Phi_{i}(\Omega)\rangle$:
\[
  |\Phi_{i}(\Omega+\mathrm{d}\Omega)\rangle = 
  P(\Omega+\mathrm{d}\Omega) |\Phi_{i}(\Omega)\rangle,
\]
which, after multiplication by $\langle\Phi_{j}(\Omega+\mathrm{d}\Omega)|$, we write in the form
\[
  \langle \Phi_i(\Omega)| \frac{\mathrm{d}}{\mathrm{d}\Omega} | \Phi_j(\Omega) \rangle = 0.
\]
The associated initial condition follows from substituting $\Omega = 0$ into Eq.~(\ref{eq:eigenvecParTransport}), giving $|\Phi_{1,2}(0)\rangle = (|f0\rangle_D \pm |g1\rangle_D)/\sqrt{2}$. Therefore, the two equal superposition states have to be eigenstates of the non-driven Hamiltonian. This is by definition also true for $|f0\rangle_D$ and $|g1\rangle_D$. The only way the two distinct pairs of vectors can be eigenstates at the same time is if the subspace they are spanning is degenerate. This can be achieved by choosing the correct frequency of the rotating frame, giving us a condition for the drive frequency at $\Omega = 0$.

For $i=j$, the differential equation above can be satisfied simply by choosing the correct phase of the eigenstates $|\Phi_{1,2}(\Omega)\rangle$. After expressing the derivative of the eigenstate in terms of the derivative of the Hamiltonian, the remaining equations for $i\neq j$ are equivalent to
\[
  \langle \Phi_1(\Omega) | \frac{\mathrm{d}H(\Omega)}{\mathrm{d}\Omega} | \Phi_2(\Omega) \rangle = 0.
\]

Now we consider that the Hamiltonian depends on $\Omega$ not only directly but also through the amplitude-dependent drive frequency $\omega_d(\Omega)$. The total derivative with respect to $\Omega$ can then be expressed using the chain-rule, leading to the equation
\begin{align}
  &\langle \Phi_1(\Omega,\omega_d) | \frac{\partial H(\Omega,\omega_d)}{\partial\Omega} | \Phi_2(\Omega,\omega_d) \rangle +
  \nonumber\\
  &\frac{\mathrm{d}\omega_d(\Omega)}{\mathrm{d}\Omega}
  \langle \Phi_1(\Omega,\omega_d) | \frac{\partial H(\Omega,\omega_d)}{\partial\omega_d} | \Phi_2(\Omega,\omega_d) \rangle
  = 0.\label{eq:diffEqnStark}
\end{align}

We can solve this differential equation for $\omega_d(\Omega)$ numerically to find the amplitude-dependent drive frequency which yields an effective Hamiltonian of the form shown in Eq.~(\ref{eq:heffForm}). 

The simple procedure for finding the solution is summarized here:

\begin{enumerate}
\item{Start with $\Omega=0$. Find $\omega_d$ by requiring $|f0\rangle_D$ and $|g1\rangle_D$ to be degenerate.}
\item{Find the eigenstates $|\Phi_{1,2}(\Omega,\omega_d)\rangle$ of the Hamiltonian $H(\Omega,\omega_d)$. For the first step when $\Omega=0$ and the eigenstates are degenerate, choose $|\Phi_{1,2}(\Omega,\omega_d)\rangle = (|f0\rangle_D \pm |g1\rangle_D)/\sqrt{2}$.}
\item{Use Eq.~(\ref{eq:diffEqnStark}) to calculate $\omega'_d(\Omega):=\mathrm{d}\omega_d(\Omega)/\mathrm{d}\Omega$}
\item{Set $\omega_d\to\omega_d+ \omega'_d(\Omega)\Delta\Omega$ and $\Omega\to\Omega+\Delta\Omega$.}
\item{Go to step 2.}
\end{enumerate}

Once the solution is known, the effective coupling $\tilde{g}(\Omega)$ is calculated from the eigenenergies $E_{1,2}(\Omega,\omega_d)$ of the two eigenstates $| \Phi_{1,2}(\Omega,\omega_d) \rangle$. Inspection of Eq.~(\ref{eq:heffForm}) shows that these eigenenergies are equal to $E_{\mathrm{offset}}\pm \tilde{g}$ and therefore
\[
  \tilde{g}(\Omega) = \frac{E_{1}(\Omega,\omega_d)-E_{2}(\Omega,\omega_d)}{2}.
\]

\section{Calculation of $\alpha_{\pm}$ and $\beta{\pm}$}
\label{app:Nonad}

In this appendix we use perturbation theory to calculate the overlaps $\alpha_{\pm}$, $\beta_{\pm}$ of the driven Hamiltonian eigenstates $\ket{\Phi_{\pm}}$ corresponding to the undriven eigenstates $\ket{\Phi_{\pm}^{(0)}} = (|f0\rangle_{D} + |g1\rangle_{D})/\sqrt{2}$. The result is then used to derive Eq.~(\ref{STIFID}). We retain only terms up to second order in the drive strength $\Omega$ and zeroth order in the Jaynes-Cummings coupling $g$. In this approximation, we can treat states with different numbers of photons as decoupled and replace the dressed states $|ij\rangle_D$ by the corresponding bare states $|ij\rangle$. The overlaps $\alpha_{\pm} = \langle\Phi_{\pm}|f0\rangle$ and $\beta_{\pm} = \langle\Phi_{\pm}|g1\rangle$ which we wish to calculate are given by
\begin{align*}
  \alpha_{\pm} &= \frac{1}{\sqrt{2}}\left(
  1 + \langle \tilde{f0}^{(2)}| f0\rangle
  \right) \\
  &= \frac{1}{\sqrt{2}}\left(
  1 - \frac{1}{2}|\langle \tilde{f0}^{(1)}| \tilde{f0}^{(1)} \rangle|^2
  \right), \\
  \beta_{\pm} &= \pm \frac{1}{\sqrt{2}}\left(
  1 + \langle \tilde{g1}^{(2)}| g1\rangle
  \right) \\
  & = \pm \frac{1}{\sqrt{2}}\left(
  1 - \frac{1}{2}\langle \tilde{g1}^{(1)}| \tilde{g1}^{(1)}\rangle
  \right),
\end{align*}
where $|\tilde{ij}^{(k)}\rangle$ are the $k$-th order corrections to the undriven eigenstates $|ij\rangle$. Specifically,
\begin{align*}
|\tilde{f0}^{(1)}\rangle &= \frac{\sqrt{3}\Omega/2}{\Delta+\alpha}|h0\rangle
-\frac{\sqrt{2}\Omega/2}{\Delta}|e0\rangle,\\
|\tilde{g1}^{(1)}\rangle &= \frac{\Omega/2}{\Delta+\alpha}|e1\rangle,
\end{align*}
which results in the following expressions for $\alpha_{\pm}$ and $\beta_{\pm}$:

\begin{eqnarray}
\nonumber \alpha_{\pm} &=& \frac{1}{\sqrt{2}}\left(1-\frac{1}{2}\left(\frac{\sqrt{2}\Omega/2}{\Delta}\right)^{2}-\frac{1}{2}\left(\frac{\sqrt{3}\Omega/2}{\Delta-\alpha}\right)^{2}\right)\\
\nonumber \beta_{\pm} &=& \pm\frac{1}{\sqrt{2}}\left(1-\frac{1}{2}\left(\frac{\Omega/2}{\Delta+\alpha}\right)^{2}\right).\\
\end{eqnarray}

\end{document}